\documentstyle[12pt,psfig]{article}

\begin{document}
\baselineskip 0.7cm

\newcommand{\gsim}{ \mathop{}_{\textstyle \sim}^{\textstyle >} }
\newcommand{\lsim}{ \mathop{}_{\textstyle \sim}^{\textstyle <} }
\newcommand{\vev}[1]{ \left\langle {#1} \right\rangle }
\newcommand{\lsp}{ \left ( }
\newcommand{\rsp}{ \right ) }
\newcommand{\lmp}{ \left \{ }
\newcommand{\rmp}{ \right \} }
\newcommand{\llp}{ \left [ }
\newcommand{\rlp}{ \right ] }
\newcommand{\labs}{ \left | }
\newcommand{\rabs}{ \right | }
\newcommand{\EV} { {\rm eV} }
\newcommand{\KEV}{ {\rm keV} }
\newcommand{\MEV}{ {\rm MeV} }
\newcommand{\GEV}{ {\rm GeV} }
\newcommand{\TEV}{ {\rm TeV} }
\newcommand{\YR}{ {\rm yr} }
\newcommand{\mgut}{M_{GUT}}
\newcommand{\mint}{M_{I}}
\newcommand{\mgra}{M_{3/2}}
\newcommand{\mll}{m_{\tilde{l}L}^{2}}
\newcommand{\mdr}{m_{\tilde{d}R}^{2}}
\newcommand{\mllXX}[1]{m_{\tilde{l}L , {#1}}^{2}}
\newcommand{\mdrXX}[1]{m_{\tilde{d}R , {#1}}^{2}}
\newcommand{\mgy}{m_{G1}}
\newcommand{\mgl}{m_{G2}}
\newcommand{\mgc}{m_{G3}}
\newcommand{\nuR}{\nu_{R}}
\newcommand{\slL}{\tilde{l}_{L}}
\newcommand{\slLi}{\tilde{l}_{Li}}
\newcommand{\sdR}{\tilde{d}_{R}}
\newcommand{\sdRi}{\tilde{d}_{Ri}}
\newcommand{\e}{{\rm e}}
\newcommand{\bsub}{\begin{subequations}}
\newcommand{\esub}{\end{subequations}}
\newcommand{\wt}{\widetilde}
\newcommand{\btable}{\begin{table}[htbp]\begin{center}}
\newcommand{\etable}[1]{ \end{tabular}\caption{#1}\end{center}\end{table} }
\newcommand{\tm}{\times}
\newcommand{\ra}{\rightarrow}
\renewcommand{\thefootnote}{\fnsymbol{footnote}}
\setcounter{footnote}{1}

\makeatletter
%
%
%
%
%
\newtoks\@stequation

\def\subequations{\refstepcounter{equation}%
  \edef\@savedequation{\the\c@equation}%
  \@stequation=\expandafter{\theequation}
  \edef\@savedtheequation{\the\@stequation}
  \edef\oldtheequation{\theequation}%
  \setcounter{equation}{0}%
  \def\theequation{\oldtheequation\alph{equation}}}

\def\endsubequations{%
  \ifnum\c@equation < 2 \@warning{Only \the\c@equation\space subequation
    used in equation \@savedequation}\fi
  \setcounter{equation}{\@savedequation}%
  \@stequation=\expandafter{\@savedtheequation}%
  \edef\theequation{\the\@stequation}%
  \global\@ignoretrue}


\def\eqnarray{\stepcounter{equation}\let\@currentlabel\theequation
\global\@eqnswtrue\m@th
\global\@eqcnt\z@\tabskip\@centering\let\\\@eqncr
$$\halign to\displaywidth\bgroup\@eqnsel\hskip\@centering
     $\displaystyle\tabskip\z@{##}$&\global\@eqcnt\@ne
      \hfil$\;{##}\;$\hfil
     &\global\@eqcnt\tw@ $\displaystyle\tabskip\z@{##}$\hfil
   \tabskip\@centering&\llap{##}\tabskip\z@\cr}

\makeatother


\begin{titlepage}

\begin{flushright}
UT-02-16
\end{flushright}

\vskip 0.35cm
\begin{center}
{\large \bf A Mini-Review of \\ Constraints on Extra Dimensions \\}
\vskip 1.2cm
Yosuke Uehara$^{a)}$

\vskip 0.4cm

$^{a)}$ {\it Department of Physics, University of Tokyo, 
         Tokyo 113-0033, Japan}\\

\vskip 1.5cm

\abstract{We present a mini-review of present constraints of the large extra
dimension scenario. We show many experiments and considerations that
can constrain the fundamental scale of the large extra dimension. 
We observe that constraints come from collider 
experiments are much weaker than those of
astrophysical and cosmological considerations. When the number of
extra dimension $n$ is smaller than 4, the constraint is so strong 
that the large extra dimension scenario cannot solve hierarchy problem. 
But when $n \ge 4$ there is still possibility that it 
can solve the hierarchy problem.}

\end{center}
\end{titlepage}

\renewcommand{\thefootnote}{\arabic{footnote}}
\setcounter{footnote}{0}

\section{Introduction}

It was believed that because of the large Planck scale: 
$G_{N}^{-1/2}=2.4 \times 10^{18} \GEV$, the gravity is so weak that
the particle physics cannot explore the world of gravity. The hierarchy
between the weak scale $G_{F}^{-1/2} \sim 100 \GEV$ and the Planck scale
is the most serious problem in the Standard Model (hierarchy problem).
Conventionally, it was explained by supersymmetry. 
The successful supersymmetric $SU(5)$ gauge 
coupling unification also encourage 
the existence of low-energy supersymmetry.

But no supersymmetric particles are found so far, and thus it is
worthwhile to search for alternatives. In fact, some groups
proposed \cite{ADD,ANTONIADIS} that the weakness of
the gravity can be explained by the existence of extra dimensions.
We are trapped in '3-brane' \cite{SUNDRUM} 
of the higher-dimensional spacetime, and
only graviton can propagate the compactified extra dimensions.
Thus the overlapping between the Standard Model 
particles and gravitons becomes small, 
and the weakness of gravity can be explained.
The true fundamental scale $M_{D}$ is the weak scale itself, and
no hierarchy problem exists. It is related to the volume of
extra dimension $V_{n}$ and the appearance fundamental scale
$M_{pl}=2.4 \tm 10^{18} \GEV$ by
\begin{eqnarray}
M_{D}^2 \sim V_{n} M_{pl}^{n+2},
\label{TRANSLATION}
\end{eqnarray}
where $n$ denotes the number of extra dimensions.
We assume $V_{n} \sim R^{n}$, where $R$ is the radius of the
compactified extra dimensions. If $n=1$, $R$ is so large that the
successful Newtonian gravity is modified, and this scenario is 
excluded. But if $n \ge 2$, $R \le 1 {\rm mm}$. In this region,
the Newtonian gravity is not tested, and thus this scenario may survive.

Also, Randall and Sundrum proposed that warped spacetime can lower
the fundamental scale \cite{RS1}. But we do not consider their scenario
in this mini-review, and concentrate on the proposal of ADD. This scenario 
is called as the large extra dimension.

The very low fundamental scale $M_{D} \sim 1 \TEV$ naively
becomes a source of very rapid proton decay, unacceptable FCNC,
$K-\bar{K}, \ B-\bar{B}$ oscillations and rare decays like $\ \mu \ra e
\gamma$. But there are many proposals against them. 
Therefore it is worthwhile to consider the ADD scenario in detail.

This mini-review is aimed to summarize the currently obtained
constraints from many experiments. This mini-review is organized 
as follows: in section \ref{COLLIDER},
we show the already obtained constraints from collider experiments.
in section \ref{COSMOLOGY}, we consider the astrophysical and
cosmological constraints on the ADD scenario. 
And in section \ref{CONCLUSION} we summarize.

{\bf Please note that, many experimental groups use different notations on
the fundamental scale, and many theorists also use different notations.
I will explain order by order what kind of notation is used, and
obtain the constraint in our notation, $M_{D}$.}

\section{Constraints from Collider Experiments}
\label{COLLIDER}

Up to now, LEP, Tevatron and HELA give lower bound on the fundamental 
scale $M_{D}$. We discuss the detail of their result in this section.
There is another review about this issue \cite{CHEUNG}.

\subsection{Real Graviton Emission}

Gravitons only weekly interact with other particles. Their interactions
are suppressed by $1/M_{pl}^{2}$. But the huge number of Kaluza-Klein
graviton modes enable us to observe graviton emission process,
like $e^{+} e^{-} \ra \gamma/Z + G$. The graviton Kaluza-Klein modes have
masses equal to $|n|/R$ and therefore the different excitations have
the mass splittings \cite{GIUDICE-RATTAZZI-WELLS} :
\begin{eqnarray}
\Delta m \sim \frac{1}{R} \sim \left( \frac{M_{D}}{\TEV} \right)^{\frac{n+2}{2}} 10^{\frac{12n-31}{n}} \ \EV.
\end{eqnarray}
The enormous number of Kaluza-Klain modes enable us to detect
the massive graviton emission processes. The number of modes with Kaluza-
Klein index between $|n|$ and $|n|+dn$ is :
\begin{eqnarray}
dN=S_{n-1} |n|^{n-1} dn, \ \ S_{n-1} = \frac{2 \pi^{n/2}}{\Gamma(n/2)},
\end{eqnarray}
where $S_{n-1}$ is the surface of a unit radius sphere in n dimensions.
From Eq. (\ref{TRANSLATION}) and $m=|n|/R$, we obtain
\begin{eqnarray}
dN=S_{n-1} \frac{M_{pl}^{2}}{M_{D}^{2+n}} m^{n-1} dm.
\end{eqnarray}
This numerator factor $M_{pl}^{2}$ is the resultant of
the enormous number of Kaluza-Klein modes, 
and it completely cancels the suppression
factor of real graviton emission, $1/M_{pl}^{2}$. 

L3 \cite{L3,L3-RESULT} and H1 \cite{H1} 
searched for this process. \cite{L3-RESULT} 
set the stringent bound on the fundamental scale. 

Here we have to make clear the notation in \cite{L3-RESULT}. 
They used that of \cite{GIUDICE-RATTAZZI-WELLS}. 
\begin{eqnarray}
G_{N}^{-1} = 8 \pi R^{n} M_{D}^{2+n} = R^{n} ((8 \pi)^{1/(2+n)} M_{D})^{2+n}.
\end{eqnarray}
So we have to multiply $(8 \pi)^{1/(2+n)}$ to the result of \cite{L3-RESULT}.
They are shown in table \ref{REAL}.

\btable
\begin{tabular}{|c|c|c|c|c|c|c|}
\hline
n & 2 & 3 & 4 & 5 & 6 & 7 \\
\hline
$M_{D} \ (\TEV)$ & 2.3 & 1.5 & 1.2 & 0.91 & 0.76 & 0.65 \\
\hline
\etable{The obtained lower bound on the fundamental scale $M_{D}$ from real
 graviton emission. \label{REAL}}

\subsection{Virtual Graviton Exchange}

Now we want to study the virtual graviton exchange processes.
We concentrate on s-channel processes, but t- and u- channel exchange
processes are completely analogous. The scattering amplitude is 
\cite{GIUDICE-RATTAZZI-WELLS}:
\bsub
\begin{eqnarray}
A &=& {\cal S} (s) \cal{T}, \\
{\cal S}(s) &\equiv& \frac{1}{M_{pl}^{2}} \sum_{n} \frac{1}{s-m_{n}^{2}}, \\
{\cal T} &\equiv& T_{\mu \nu} T^{\mu \nu} - \frac{1}{n+2} T_{\mu}^{\mu} T_{\nu}^{\nu}.
\end{eqnarray}
\esub
Again, this amplitude is suppressed by $M_{pl}$. But enormous
number of Kaluza-Klein modes enables us to investigate the virtual graviton
exchange processes by $e^{+} e^{-} \ra \gamma \gamma, \ e^{+} e^{-} \ra
\mu^{+} \mu^{-},$ etc.

ALEPH \cite{ALEPH}, DELPHI \cite{DELPHI}, L3 \cite{L3}, 
OPAL \cite{OPAL}, H1 \cite{H1} and D0 \cite{D0} searched for
this virtual graviton exchange processes. Among them, the constraint
from D0 collaboration is the most stringent. 
Instead of our fundamental scale $M_{D}$, D0 used the cutoff scale
$M_{S}$ defined as \cite{Han-Lykken-Zhang}:
\begin{eqnarray}
\kappa^{2} R^{n} = 8 \pi (4 \pi)^{(n+2)} \Gamma(n/2) M_{S}^{-(n+2)}, \label{MSeq}
\end{eqnarray}
where $\kappa = \sqrt{16 \pi G_{N}^{(n+4)}} \hat{\kappa}$.
The obtained bound on $M_{S}$ are shown in \cite{D0}. From these
values, we obtain the bound on the fundamental value $M_{D}$, defined
by us. They are shown in table \ref{VIRTUAL}.

\btable
\begin{tabular}{|c|c|c|c|c|c|c|}
\hline
n & 2 & 3 & 4 & 5 & 6 & 7 \\
\hline
$M_{D} \ (\TEV)$ & 0.88 & 0.77 & 0.58 & 0.47 & 0.39 & 0.35 \\
\hline
\etable{The obtained lower bound on the fundamental scale
 $M_{D}$ from virtual graviton exchange. \label{VIRTUAL}}

\subsection{Universal Torsion-Induced Interaction}

Even in the minimal Kaluza-Klein scenario wherein only gravity
exists in the bulk while all standard model fields are localized
on a four-dimensional brane, fermions always induce antisymmetric
pieces, or torsion, in the gravity connection \cite{LEBEDEV1}.
They affect Z-pole electroweak observables. In order not to
destroy the success of the SM Z-pole experiments, they derived
$M_{S} > 28 \TEV \ (n=2)$ at $3- \sigma$ level. $M_{S}$ is defined
in (\ref{MSeq}) and This constraint becomes
\begin{eqnarray}
M_{D} > 18 \TEV \ (n=2).
\end{eqnarray}
This is the strongest collider constraint. They also considered
the case of Randall-Sundrum scenario \cite{LEVEDEV2}.

\section{Constraints from Astrophysical and Cosmological Considerations}
\label{COSMOLOGY}

If we consider the large extra dimension scenario, there are
Kaluza-Klein graviton modes in the universe. 
They affect the evolution of supernovas, neutron stars,
cosmic diffuse gamma rays, and lead to early matter domination.
And furthermore, the ultra-high energy cosmic rays can create
black holes and may be detected by cosmic ray detectors.
From these considerations, we can set very strong bound on
the fundamental scale $M_{D}$. 

\subsection{Supernova and Neutron Star}

The existence of massive Kaluza-Klein gravitons affects the
phenomenology of SN1987A and neutron star.

SN1987A lose energy by Kaluza-Klein graviton emission.
such gravitons are created by $\gamma \gamma \ra G G, \ e^{+} e^{-} \ra
G G, \ e \gamma \ra e G$, gravi-bremsstrahlung and 
nucleon-nucleon bremsstrahlung. Therefore
in order not to lose too much energy by graviton emission, 
the fundamental scale should not be large. 

The references in \cite{SN1987A} use different definition of
the fundamental scale, and we rearranged their result
into our defined value $M_{D}$. Then the result becomes:
\bsub
\begin{eqnarray}
M_{D} \ge 26.5-188 \ \TEV \ (n=2), \\
M_{D} \ge 2.1-13 \ \TEV \ (n=3).
\end{eqnarray}
\esub
The obtained bounds depend on the supernova temperature.

Next, we consider neutron stars. The Kaluza-Klein gravitons around
neutron stars decay into photons, electrons, positrons, and neutrinos.
They hit the neutron stars and heat them.
The requirement that neutron stars are not excessively heated by
Kaluza-Klain graviton decay products implies \cite{HANNESTAD-RAFFELT} :
\bsub
\begin{eqnarray}
M_{D} &\ge& 1100 - 3600 \ \TEV \ (n=2), \\
M_{D} &\ge& 19 - 190 \ \TEV \ (n=3).
\end{eqnarray}
\esub
Note that \cite{HANNESTAD-RAFFELT} use $M_{pl}=1.2 \tm 10^{19} \GEV$,
which do not agree with our definition $M_{pl}=2.4 \tm 10^{18} \GEV$.
We take into account this difference.

\subsection{Cosmic Diffuse Gamma Ray}

In standard cosmology, big bang nucleosynthesis (BBN) provides a
detailed and accurate understanding of the observed light element
abundances. But the existence of massive Kaluza-Klein gravitons
can destroy BBN completely. we must set ``normalcy temperature''
$T_{*}$, when the extra dimension are virtually empty of energy density.
$T_{*}$ is about $1 \MEV$, and the result highly depend on the
value of $T_{*}$.

After successful BBN, Kaluza-Klein gravitons are produced through the
process $\nu \bar{\nu} \ra G$, for example. And when it decay into
two photons, it can destroy the observed cosmic diffuse gamma ray 
background. It have been measured in the $800 \KEV$ to $30 \MEV$ 
energy range, and from the experiment we can set \cite{HALL-SMITH} :
\bsub
\begin{eqnarray}
M_{D} &\ge& 210-660 \ \TEV \ (n=2), \\
M_{D} &\ge& 8.3-22.9 \ \TEV \ (n=3).
\end{eqnarray}
\esub
In deriving these constraints we consider the definition
difference of the fundamental scale.

\subsection{Early Matter Domination}

Again, Kaluza-Klein massive gravitons are created by
gravi-bremsstrahlung etc. Such massive gravitons inject
extra massive matter in the universe, and lead to a more rapid
decline in the CMB temperature. The increased cooling rate means
that by the time the CMB has cooled to 2.73K the universe is
still be much too young to hold the objects we observe in ours.
From this fact, we can set lower bound on fundamental scale.
The energy density of massive gravitons highly depends on
the QCD scale, and so the obtained bound also depend on it. 
The bound is \cite{FAIRBAIRN} :
\bsub
\begin{eqnarray}
M_{D} &\ge& 161-1900 \ \TEV \ (n=2), \\
M_{D} &\ge& 12-98 \ \TEV \ (n=3), \\
M_{D} &\ge& 2.3-14 \ \TEV \ (n=4).
\end{eqnarray}
\esub

\subsection{Can Black Hole Production by Cosmic Rays Constrain the
Fundamental Scale?}

If the large extra dimension scenario is true, black holes are created
when the collision energy of two particles is larger than
the fundamental scale. Ultra-high energy cosmic rays provide the most
promising window to observe black holes or p-branes before LHC starts
\cite{COSMICRAY}. Cosmic rays penetrate the atmosphere of the earth
and collide with some nuclei, producing black holes. They
immediately evaporate and lead to energetic showers. There exists a
paper which claims that AGASA already 
provides the most stringent bound on the fundamental scale for $n \ge 5$
\cite{AGASALIMIT}.

Of course, the black hole production cross section suffers from many
uncertainties, say grey body factors, angular momentum effects and
finite impact parameters. Furthermore if the mass of black hole 
is near $M_{D}$, quantum gravity effects may drastically change
the cross section. 

But these uncertainties are under control \cite{Feng}.
First consider the grey body factors. these are not important for the bound 
if one does not see events, the exact appearance of the predicted 
events is not important (other than the fact that these events are 
visible, but we do not know any reason to expect black holes
decay only to neutrinos, gravitons, and muons).
Next, while angular momentum is important, all analyses 
incorporating this lead to small effects.  
And finally, while quantum effects are certainly important
at $M_{BH} \sim M_{D}$, one can suppress 
this dependence by taking $M_{BH,min}$ to be a few $M_{D}$.

So I use the limit obtained from AGASA. 
Conservative bounds are \cite{AGASALIMIT}:
\bsub
\begin{eqnarray}
1.1 \TEV < &M_{D}& < 1.2 \TEV \ (n=5), \\
1.1 \TEV < &M_{D}& < 1.3 \TEV \ (n=6), \\
1.2 \TEV < &M_{D}& < 1.3 \TEV \ (n=7). 
\end{eqnarray}
\esub

\section{Summary}
\label{CONCLUSION}

To summarize, we have shown the currently 
obtained lower bound on the fundamental
scale of the large extra dimension. It is summarized in
table \ref{SUMMARY}. You notice that 
the constraints from present colliders are so
weak that astrophysical and cosmological ones dominate.
If $n \le 3$, the constraints from astrophysics and cosmology are
so stringent that we cannot solve the hierarchy problem. But if
$n \ge 4$, the constraints are very weak and there is still possibility
that the large extra dimension is the solution of the hierarchy problem.

\btable
\begin{tabular}{|c|c|c|c|}
\hline
n & 2 & 3 & 4 \\
\hline
$M_{D} \ (\TEV)$ & 1100 (neutron star) & 19 (neutron star) & 2.3 (early
 matter domination) \\
\hline
n & 5 & 6 & 7 \\
\hline
$M_{D} \ (\TEV)$ & 1.1 (AGASA) & 1.1 (AGASA) & 1.2 (AGASA) \\
\hline
\etable{The most conservative bound on the fundamental scale of the 
large extra dimension. \label{SUMMARY}}

\section*{Note Added}

After the completion of this letter, we were informed many articles
which were concerned with this letter. First, we learned that Tevatron
can put lower bound on the fundamental scale, $M_{D} > 1.4 \TEV$ 
\cite{HOFFMANN}. Next, we noticed that a simple extension of the
original ADD scenario can escape from the bounds described in
this article \cite{STOJKOVIC}. Finally our attention were drawn
to \cite{GIUDICE}, which examines model-independent test of 
extra-dimensional gravity performed at high-energy colliders
using transplanckian collisions.

\section*{Acknowledgement}

We thank H.~Takayanagi for stimulating discussions.
Y.U. thank Japan Society for the 
Promotion of Science for financial support.

\end{document}